\documentstyle[12pt]{article}
\def\be{\begin{equation}}
\def\ee{\end{equation}}
\def\ba{\begin{eqnarray}}
\def\ea{\end{eqnarray}}
\def\d{\partial}
\def\a{\alpha}
\def\th{\theta}
\def\tt{\theta_{12}}
\def\tp{\theta_{12}^+}
\def\tm{\theta_{12}^-}
\def\tpa{\theta_1^+}
\def\tpb{\theta_2^+}
\def\tma{\theta_1^-}
\def\tmb{\theta_2^-}
\def\g{\gamma}
\def\z{z_{12}}
\def\Gb{\bar{G}}
\def\AA{{\cal{A}}}
\def\BB{{\cal{B}}}
\def\II{{\cal{I}}}
\def\TT{{\cal{T}}}
\def\tb{\bar{\theta}}
\def\L{\Delta}
\def\P{\Phi}

\def\Oa{O_{G=1}}
\def\Ob{O_{G=2}}
\def\Obt{\tilde O_{G=2}}
\def\Oc{O_{G=3}}

\begin{document}
\renewcommand{\theequation}{\thesection.\arabic{equation}}
\newcommand{\beq}{\begin{equation}}
\newcommand{\eeq}[1]{\label{#1}\end{equation}}
\newcommand{\ber}{\begin{eqnarray}}
\newcommand{\eer}[1]{\label{#1}\end{eqnarray}}
\begin{center}
        February, 1993          \hfill    NSF-ITP-93-17\\
			        \hfill    ITP-SB-93-09\\
			        \hfill    RIP-148-93\\
                                \hfill    hep-th/9302049\\

\vskip .5in

{\large \bf On the BRST Operator Structure of the $N=2$ String}
\vskip .5in

{\bf Amit Giveon} \footnotemark \\

\footnotetext{e-mail address: GIVEON@HUJIVMS.bitnet}

\vskip .1in

{\em Racah Institute of Physics, The Hebrew University\\
  Jerusalem, 91904, ISRAEL} \\

\vskip .15in

       and

\vskip .15in

{\bf Martin Ro\v cek} \footnotemark \\

\footnotetext{e-mail address: rocek@insti.physics.sunysb.edu}

\vskip .1in

{\em Institute for Theoretical Physics \\
State University of New York at Stony Brook \\
Stony Brook, NY 11794-3840 USA}\\
\vskip .1in
\end{center}
\vskip .4in
\begin{center} {\bf ABSTRACT } \end{center}
\begin{quotation}\noindent
The BRST operator cohomology of $N=2$ $2d$ supergravity coupled to matter
is presented. Descent equations for primary superfields of the matter
sector are derived. We find one copy of the cohomology at ghost number one,
two independent copies at ghost number two, and conjecture that there is a
copy at ghost number three. The $N=2$ string has a twisted $N=4$
superconformal symmetry generated by  the $N=2$ superstress tensor, the
BRST  supercurrent, the antighost superfield, and the ghost number
supercurrent.
\end{quotation}
\vfill
\eject
\def\baselinestretch{1.2}
\baselineskip 16 pt
\noindent
\section{Introduction}
\setcounter{equation}{0}

The BRST cohomology of a string theory describes the physical states of
the theory (for a review, see \cite{GSW});  moreover, the explicit
understanding of the cohomology in a covariant gauge has been crucial for
the construction of both $N=0$ and $N=1$ string field theories.  In
nontrivial backgrounds, the BRST cohomology has revealed a deep structure
related to the underlying symmetries of string theory (for example, see
\cite{W}).

The $N=2$ string is a relatively simple string theory (for a review, see
for example \cite{OV}). It might therefore be practical to first understand
aspects of the $N=2$ string (for example, underlying symmetries are studied
in \cite{givshap}), and to use the insight gained to learn about more
realistic, and more complicated, string theories.

In general, there is an equivalence between the BRST cohomology on states
in Fock space, that is states annihilated by the BRST charge $Q$ that are
not themselves $Q$ of another state, and BRST cohomology on operators, that
is operators that (super)commute with $Q$ but are not themselves
(super)commutators with $Q$. However, only operator cohomology can be
conveniently described in superspace.\footnote{There are certain special
operators, such as the picture changing operator for $N=1,2$ \cite{FMS}
and the instanton number operator for $N=2$ \cite{NB} that have not been
constructed in superspace, and that we will not discuss.}

The state cohomology for the $N=2$ string in ($2,2$)-dimensional flat
Minkowski space was constructed in \cite{B}.  Here we consider the operator
$N=2$  BRST cohomology in an arbitrary background.

More generally, in this paper we explore aspects of the BRST operator
structure of the $N=2$ string. We present the operator cohomology in
superspace, and derive descent equations for primary superfields.  We also
find that the $N=2$ string has a twisted critical $N=4$ superconformal
symmetry generated by the $N=2$ superstress tensor, the BRST supercurrent,
the antighost superfield,  and the ghost number supercurrent.

The paper is organized as follows:  In section 2, we review BRST operator
cohomology for $N=0,1$ strings. In section 3, we derive analogous formulae
for $N=2$ strings: given an $N=2$ primary with vanishing dimension and
charge, $h=q=0$, we construct corresponding operators in the BRST
cohomology, one at ghost number $G=1$, two independent operators at $G=2$,
and a fourth operator at $G=3$ that we conjecture is physical; all are
described in superspace, without bosonization, etc. We derive the $N=2$
descent equations, which explain how the operators arise.  We find that a
certain conjugation relation plays the role that is played by derivatives
for the $N=0,1$ cases. We also consider the explicit example of the $N=2$
string in a toroidal background, and find the usual discrete states. In
section 4, we show how the superstress tensor $T$, the BRST supercurrent
$j$, the antighost superfield $B$, and the ghost number supercurrent $BC$
form a twisted $N=4$ algebra.  In section 5 we end with some discussion and
comments. Finally, in the Appendix, we give some computational details.

\section{Review of BRST Cohomology for $N=0,1$ strings}
\setcounter{equation}{0}

We first review  the BRST cohomology of the $N=0$ bosonic string (see ref.
\cite{GSW}). The stress tensor $T(z)=T^{\P}+T^{bc}$ combines the matter
sector ($T^{\P}$) with the reparametrization $b,c$ ghost system ($T^{bc}$).
The stress tensor acts on a primary field of the matter sector $\P$ with
dimension $h$ via the operator product expansion (OPE)
\be
T_1^{\P}\P_2\sim \frac{h}{\z^2}\P_2 + \frac{1}{\z}(\d\P )_2 + ...\, ,
\label{TP0}
\ee
where the dots in the OPE indicate nonsingular terms, and we use the
notation $f_1\equiv f(z_1)$, and $\z=z_1-z_2$. The ghosts $c$ and $b$ have
dimensions $-1$ and $2$ with respect to the ghosts stress tensor
$T^{bc}=c\d b + 2(\d c)b$. They satisfy
\be
c_1b_2\sim b_1c_2\sim \frac{1}{\z}+...\, .
\ee
The $N=0$ BRST current is
\be
j=c(T^{\P}+\frac{1}{2}T^{bc})+\frac{3}{2}\d^2 c,
\label{j0}
\ee
where the total derivative is chosen such that $j$ is a primary field with
$h=1$. At criticality, the nilpotent BRST charge
\be
Q=\frac{1}{2\pi i}\oint dz\; j(z)
\label{Q0}
\ee
acts on a field $O$ by
\be
Q(O)=[Q,O(z_2)\}\equiv\frac{1}{2\pi i}
\oint_{C_{z_2}} dz_1\; j(z_1)O(z_2),
\label{QO0}
\ee
where the contour $C_{z_2}$ surrounds $z_2$ once.

Next we discuss BRST cohomology classes. For a primary field of the matter
sector $\P$ with $h=1$, the fields
\be
O_{G=1}=c\P, \qquad  O_{G=2}=c\d c\P
\label{O12}
\ee
are BRST invariant, namely, $Q(O_{G=1})=Q(O_{G=2})=0$. The operators
$O_{G=1}$ and $O_{G=2}$ are not $Q$-anticommutators, and therefore, they
represent cohomology classes. The label $G$ in (\ref{O12}) describes the
ghost number of the operator $O$ (with the convention $G(c)=1$, $G(b)=-1$).
The two operators in (\ref{O12}) generate  the same physical state in
different ghost vacua.

An element of the BRST cohomology $O_{G=1}$  is related to its
corresponding primary field $\P$ by the descent equation
\be
Q(\P )=\d O_{G=1}.
\label{desc0}
\ee
Moreover, $O_{G=1}$ and $O_{G=2}$ are related via the equation
\be
Q(c\P_h)=(1-h)c\d c \P_h,
\label{cPcdcP}
\ee
where $\P_h$ is off-shell when $h\neq 1$. Using equation (\ref{cPcdcP})
(and continuity in $h$) in the on-shell limit $h\rightarrow 1$, one finds
that $O_{G=2}$ is physical. $\Ob$ also satisfies the descent equation
\be
Q(\d c\P)=\d \Ob .
\ee

The critical bosonic string can (often) be regarded as a topological
conformal field theory \cite{DVV}. The OPE's of the operators $T,b,j$ and
$J=bc$ are `almost' closed on a twisted $N=2$ superconformal algebra
(closing the algebra requires two more generators: $c$ and $c\d c$). The
untwisted  $N=2$ superconformal algebra is given by twisting $T$ to
$T+\frac{1}{2}\d J$.\footnote{ If the background has a scalar field, then
it is possible to add total derivative terms to $j$ and $J$, so that the
modified generators $T,b,j,J$ form a twisted $N=2$ algebra which closes
without adding more generators~\cite{GRS}.} This concludes our discussion
of the $N=0$ case.

Next we discuss the BRST cohomology of the $N=1$ superstring in the NS
sector.  We first set the notation \cite{FMS}. We define
\ba
z_{12}&=&z_1-z_2-{\th}_1{\th}_2,\nonumber\\
\th_{12}&=&\th_1-\th_2,
\label{z12}
\ea
and the superderivative
\be
D=\frac{\d}{\d\th}+\th\d,\qquad D^2=\d,
\label{par}
\ee
where $\d\equiv \frac{\d}{\d z}$.
A function $f_1\equiv f(z_1,\th_1)$ can be expanded around $z_2,\th_2$:
\be
f_1=f_2+\tt(Df)_2+\z(\d f)_2+\tt\z(D\d f)_2+...
\label{f1}
\ee

An $N=1$ stress tensor $T=G+\th T_B$, where $T_B$ is the bosonic stress
tensor, obeys the $N=1$ superconformal algebra expressed via the operator
product expansion
\be
T_1T_2\sim \frac{\frac{3}{8}c}{\z^3}+\frac{\frac{3}{2}\th_{12}}{\z^2}T_2
+\frac{\frac{1}{2}}{\z}(D T)_2+\frac{\th_{12}}{\z}(\d T)_2+...\; .
\ee
The $N=1$ $2d$ theory combines a matter sector and a ghost sector. The
$N=1$ energy-momentum tensor of the matter sector $T^{\P}$ acts on a
primary superfield of the matter sector $\P$ with dimension $h$ as
\be
T^{\P}_1\P_2\sim
\frac{h\tt}{\z^2}\P_2+\frac{\frac{1}{2}}{\z}(D\P)_2
+\frac{\th_{12}}{\z}(\d\P)_2+...
\label{TP1}
\ee
The superghosts $C$ and $B$ have dimensions $-1$ and $3/2$ with respect to
the superghost stress tensor $T^{BC}=-C(\d B)+\frac{1}{2}DCDB-\frac{3}{2}
(\d C)B$. They satisfy
\be
C_1 B_2\sim B_1 C_2\sim\frac{\tt}{\z}+... .
\label{BC}
\ee
The N=1 BRST current is
\be
j=C(T^{\P}+\frac{1}{2}T^{BC})-\frac{3}{4}D(C(DC)B),
\label{j1}
\ee
The total derivative term in
(\ref{j1}) is chosen such that $j$ is a primary superfield with $h=1/2$. At
criticality, the nilpotent BRST charge
\be
Q=\frac{1}{2\pi i}\oint dz d\th\; j(z,\th),
\label{Q1}
\ee
acts on a superfield $O$ by
\be
Q(O)=[Q,O(z_2,\th_2)\}\equiv\frac{1}{2\pi i}
\oint_{C_{z_2}} dz_1 d\th_1\; j(z_1,\th_1)O(z_2,\th_2),
\label{QO1}
\ee
where the contour $C_{z_2}$ surrounds $z_2$ once.

Next we describe BRST cohomology classes. For a primary superfield of the
matter sector $\P$ of dimension $h=1/2$, the combinations
\be
O_{G=1}=CD\P-\frac{1}{2}(DC)\P
\label{O}
\ee
and
\be
O_{G=2}=CD(\d C\P )-\frac{1}{2}(DC)(\d C\P )
\label{O'}
\ee
are BRST invariant, namely, $Q(O_{G=1})=Q(O_{G=2})=0$.  The operators $O$
are not   $Q$-(anti)commutators, and therefore, they represent cohomology
classes. The label $G$ in (\ref{O},\ref{O'}) describes the ghost number of
the operator $O$ (with the convention $G(C)=1$, $G(B)=-1$). Unlike the
$N=0$ case, $O_{G=1}$ and $O_{G=2}$ differ  not only by a choice of their
ghost vacua, but also by a picture changing operation
\cite{FMS},\cite{erik}.

The superfields $\P$ and $O_{G=1}$ satisfy the descent equation
\be
Q(\P)=DO_{G=1}.
\label{desc1}
\ee
In terms of components one gets
\be
Q(\P_0)=O_1, \qquad Q(\P_1)=\d O_0,
\ee
where $\P=\P_0+\th\P_1$, and $O_{G=1}=O_0+\th O_1$, and thus $O_0$ is the
only physical component amongst the components of $\Oa$.
Moreover, $O_{G=1}$ and $O_{G=2}$ are related via the equation
\be
Q(O_{G=1}^h)=(\frac{1}{2}-h)O_{G=2}^h,
\label{O12N1}
\ee
where $O^h$ is constructed from an off-shell $\P_h$ when $h\neq 1/2$.
Using equation (\ref{O12N1}) (and continuity in $h$) in the on-shell limit
$h\rightarrow 1/2$, one finds that $O_{G=2}$ is physical.
$\Ob$ also satisfies the descent equation
\be
Q(\d C\P)=D\Ob .
\ee

This concludes the review of the $N=1$ BRST case. In the next section we
discuss the $N=2$ BRST case in $N=2$ superspace,  the BRST invariant
operators, and the descent equation.

\section{BRST Cohomology of the $N=2$ String}
\setcounter{equation}{0}

We first set the notation. We define
\ba
\z&=&z_1-z_2-\frac{1}{2}(\tpa\tmb+\tma\tpb),\nonumber\\
\tt^{\pm}&=&\th_1^{\pm}-\th_2^{\pm},
\label{ztt}
\ea
and the spinor derivatives
$$
D_+=\frac{\d}{\d\th^+}+\frac{1}{2}\th^-\d, \qquad
D_-=\frac{\d}{\d\th^-}+\frac{1}{2}\th^+\d, \qquad \{D_+,D_-\}=\d,
$$
\be
D_{\pm 1}z_{12} = D_{\pm 2}z_{12}=\frac12\th_{12}^\mp ,
\label{D+-}
\ee
where $\d\equiv\frac{\d}{\d z}$. A function $f_1\equiv f(z_1,\tpa,\tma)$
can be expanded around $z_2,\tpb,\tmb$:
\ba
f_1=f_2+\tm (D_-f)_2 + \tp (D_+f)_2 + \z(\d f)_2 +
\frac{1}{2}\tp\tm([D_-,D_+]f)_2 \nonumber\\
+ \z\tm (D_-\d f)_2 + \z\tp (D_+\d f)_2
+ \frac{1}{2}\z\tp\tm(\d [D_-,D_+]f)_2 + ...\; .
\label{ff2}
\ea

We define an operator $\L^{h,q}$ to be
\be
\L^{h,q}=h\frac{\tp\tm}{\z^2}+\frac{\tp}{\z}D_+ - \frac{\tm}{\z}D_- +
\frac{\tp\tm}{\z} \d - \frac{q}{\z},
\label{L}
\ee
and for simplicity we denote
\be
\L^{h}\equiv \L^{h,0}.
\label{Lh}
\ee

An $N=2$ stress tensor  $T=J+\th^+ G_+ + \th^- G_- + \th^+\th^- T_B$,
where $T_B$ is the bosonic stress tensor, obeys the $N=2$ superconformal
algebra expressed via the OPE
\be
T_1T_2\sim \frac{\frac{1}{3}c}{\z^2} + \L^1 T_2 + ...,
\label{TT2}
\ee
where $\L^1$ is defined in (\ref{Lh},\ref{L}), and $c$ is the central
charge. The $N=2$ $2d$ theory combines a matter sector with a ghost sector.
The $N=2$ superstress tensor of the matter sector  $T^{\P}$ acts on a
primary superfield of the matter sector $\P_{h,q}$  with dimension $h$ and
$U(1)$-charge $q$ as
\be
T^{\P}_1(\P_{h,q})_2 \sim \L^{h,q}(\P_{h,q})_2+...\; .
\label{TP2}
\ee
The $N=2$ superghosts $C$ and $B$ have components
\ba
C=c+\th^+\g_++\th^-\g_- +\th^+\th^-\xi, \nonumber\\
B=\eta+\th^+\beta_+ - \th^- \beta_- +\th^+\th^- b,
\label{CB2}
\ea
and satisfy
\be
C_1B_2\sim B_1C_2\sim\frac{\tp\tm}{\z}+...\; .
\label{CBBC}
\ee
The superstress tensor of the ghost system is
\be
T^{BC}=\d (CB)-(D_+CD_-B+D_-CD_+B)
\label{TBC2}
\ee
The superghosts $C$ and $B$ have dimension and charge
$h(C)=-1,\, q(C)=0,\, h(B)=1, \, q(B)=0$, with respect to $T^{BC}$.
The total stress tensor of an $N=2$ string is
\be
T=T^{\P}+T^{BC}.
\label{TT}
\ee
At criticality, the total central charge is $0$, which implies that
the central charge of the matter sector is $c=6$.

The $N=2$ BRST current is
\ba
j&=&C(T^{\P}+\frac{1}{2}T^{BC})-\frac{1}{2}[D_-(CD_+(BC))+D_+(CD_-(BC))]
\nonumber\\
&=&CT+B(D_+CD_-C - C\d C).
\label{j2}
\ea
The total derivative terms in (\ref{j2}) are chosen such that $j$ is a
primary superfield of $T$ in (\ref{TT}) with $h=q=0$.  As for the $N=1$
case in (\ref{Q1}),  at criticality, the  nilpotent  $N=2$ BRST charge
\be
Q=\frac{1}{2\pi i}\oint dz d\th^+d\th^- j(z,\th^+,\th^-),
\label{Q2}
\ee
acts on a superfield $O$ by
\be
Q(O)=[Q,O(z_2,\tpb,\tmb)\}\equiv\frac{1}{2\pi i}
\oint_{C_{z_2}} dz_1 d\tpa d\tma j(z_1,\tpa,\tma)O(z_2,\tpb,\tmb),
\label{QO2}
\ee
where the contour $C_{z_2}$ surrounds $z_2$ once.

Next we describe the BRST cohomology classes. First we define
\be
O(\Psi )\equiv D_-(CD_+\Psi )-D_+(CD_-\Psi ).
\ee
Then for a primary superfield of
the matter sector with $h=q=0$, $\P\equiv \P_{0,0}$, the combinations
\ba
O_{G=1}&=&O(\P ),
\label{O1}\\
O_{G=2}&=&O(\d C\P ),
\label{O2}\\
\tilde O_{G=2}&=&\frac12O([D_-,D_+]C\P ),
\label{O2t}\\
O_{G=3}&=&\frac12O(\d C[D_-,D_+]C \P ),
\label{O3}
\ea
are BRST invariant, namely, $Q(\Oa)=Q(\Ob)=Q(\Obt)=Q(\Oc)=0$. The operators
$\Oa ,\;\Ob$, and $\Obt$ are not $Q$-(anti)commutators, and therefore,
represent cohomology classes. We conjecture that $\Oc$ is $Q$-nontrivial as
well. The label $G$ in (\ref{O1})-(\ref{O3}) describes the ghost number of
the operator $O$ (with the convention $G(C)=1,\, G(B)=-1$).

We have described the BRST cohomology in particular pictures; indeed,
as in the $N=1$ case, the operators $\Oa ,\;\Ob ,\;\Obt$, and $\Oc$ not only
correspond to different ghost vacua, but also differ by picture
changing operations.

The proof that $\Oa$ is BRST invariant but is not a $Q$-commutator is
straightforward. It helps to use the relations
\be
Q(C)= C\d C-D_+CD_-C,
\label{jC}
\ee
\be
Q(\P )=-D_-(CD_+\P )-D_+(CD_-\P ),
\label{jP}
\ee and
\be
Q(f(C)g(\P ))=Q(f(C))g(\P ) + (-)^f f(C)Q(g(\P )) ,
\label{jfg}
\ee
where $f$ and
$g$ are arbitrary functions of $C$ and $\P$ and their derivatives,
respectively; the last relation follows because the structure of $Q$ is
such that there are no double contractions with $f$ and $g$.

Let us denote
\be
Y_\pm\equiv CD_{\pm}\P .
\label{Ypm}
\ee
Equation (\ref{jP}) can be interpreted as the descent equation
\be
Q(\P )=-D_-Y_+ - D_+Y_- \equiv -*O_{G=1}.
\label{QP2}
\ee
The $*$ operation in (\ref{QP2}) changes
$D_-Y_+\pm D_+Y_-$ to $D_-Y_+\mp D_+Y_-$.

We recall that $\Oa$ in (\ref{O1}) is the antisymmetric combination $\Oa
=D_-Y_+-D_+Y_-$. The descent equation (\ref{QP2}) tells us that the
symmetric combination $*\Oa =D_-Y_++D_+Y_-$ is a $Q$-commutator of a
primary superfield. However, for any $Y_\pm$ that obeys $Q(D_-Y_++D_+Y_-)=
0$, one has $Q(Y_\pm )=D_\mp Y_{(\pm 2)}$, and hence the antisymmetric
combination $*(D_-Y_++D_+Y_-)=D_-Y_+-D_+Y_-$ is also annihilated by $Q$.
Consequently $O_{G=1}$ is $Q$-closed. This is analogous to the $N=0$
($N=1$) case, where the descent equation (\ref{desc0}) ((\ref{desc1}))
relates a (super)derivative of an element in the BRST cohomology to the
$Q$-(anti)commutator of its corresponding primary (super)field; here
the $*$ operation plays the role of the derivative.

Furthermore,  generically $\Oa$ cannot be $Q$-exact, since there is no
natural candidate operator $\Psi$  for $Q(\Psi )$ of the correct ghost
number and dimension other than $\P$ itself; also, on physical grounds,
since $\P$ is an $h=q=0$ primary, it creates a physical state, and hence
$\Oa$ should not be trivial. Finally, we have explicitly checked that there
are no further operators bilinear in $C$ and $\P$ in the cohomology.

Expanding the descent equation (\ref{QP2}) in components one finds
\be
Q(\P_1 ) = \frac{1}{2}\d O_0,
\label{P1}
\ee
\be
Q(\P_{\pm})=\pm O_{\pm},
\label{Ppm}
\ee
and
\be
Q(\d \P_0) = -2O_1,
\label{dP0}
\ee
where
$\P=\P_0+\th^+\P_+ +\th^-\P_- + \th^+\th^-\P_1$ and $\Oa =O_0+\th^+
O_+ +\th^- O_- + \th^+\th^- O_1$ with
\ba
O_0&=&\g_-\P_+-\g_+\P_--2c\P_1\nonumber\\
O_+&=&-\xi+\frac{1}{2}\d c-\g_+\P_1-\frac{1}{2}\g_+\d\P_0+c\d\P_+
\nonumber\\
O_-&=&-\xi-\frac{1}{2}\d c+\g_-\P_1-\frac{1}{2}\g_-\d\P_0+c\d\P_-
\nonumber\\
O_1&=&\frac{1}{2}\d(\g_-\P_++\g_+\P_--c\d\P_0).
\label{Ocom}
\ea
{}From eqs.
(\ref{Ppm},\ref{dP0}) we learn that $O_0$ is the only physical operator
amongst the components of the superfield $\Oa$; it is referred to as the
(0,0)-picture physical operator (see \cite{L} for a discussion of picture
changing for the $N=2$ string). This concludes the discussion of BRST
cohomology at ghost number one.

To prove that $O_{G=2}$ in
(\ref{O2}) is BRST invariant, we define $O_{G=2}^h$ by the relation
\be
Q(O_{G=1}^h)=-hO_{G=2}^h,
\label{O12N2}
\ee
where $O_{G=1}^h$ is defined  by replacing $\P$ in $O_{G=1}$ with an
off-shell primary superfield with zero U(1)-charge: $\P_h\equiv \P_{h,0}$.
Equation (\ref{O12N2}) is similar to the $N=0$ case in eq. (\ref{cPcdcP}),
and to the $N=1$ case in eq. (\ref{O12N1}), where $O_{G=2}$ is related to
a $Q$-anticommutator of $O_{G=1}$ at the limit $h\rightarrow 1$ and
$h\rightarrow 1/2$, respectively;  here we will find that $O_{G=2}$ in eq.
(\ref{O2}) is the limit $h\rightarrow 0$ of $O_{G=2}^h$, and thus is
$Q$-closed.

As a trick for computing $O_{G=2}^h$ we define the operator $Q_h$ by the
relations
\be
Q(\P_h)=-*O_{G=1}^h+hQ_h(\P_h),\qquad Q_h(C)=0.
\label{QQh}
\ee
Note that $*O_{G=1}^h$ has the same functional form as $*O_{G=1}$ in
(\ref{QP2}) with $\P_h$ instead of $\P$, and therefore has no explicit
$h$-dependence. Equation (\ref{QQh}) implies that the operator $Q_h$ acts
on $\P_h$ and $C$ by
\be
Q_h(\P_h)=\d C \P_h, \qquad Q_h(C)=0.
\label{QhP}
\ee
Because there is no explicit $h$ dependence in $O_{G=1}^h$, and since
$Q(O_{G=1})=0$, one gets
\be
Q(O_{G=1}^h)=hQ_h(O_{G=1}^h),
\ee
and hence from (\ref{O12N2}) it follows that
\be
Q_h(O_{G=1}^h)=-O_{G=2}^h.
\label{QhO1}
\ee
This holds in the limit $h\rightarrow 0$ as well, and thus, using
eq. (\ref{QhP}) one recovers $O_{G=2}$ in eq. (\ref{O2}).

To verify that $\Ob$ is {\it not} $Q$-exact we have checked that acting
with $Q$ on the 6 independent bilinear terms in $C$ and $\P$, it is
impossible to get $\Ob$ (see Appendix). We can also  show that $*\Ob$ is
$Q$-exact, which implies that we can construct $\Ob$ from a descent
equation
\be
Q(\d C\P)=-*\Ob .
\ee
This is a consequence of the equalities
\be
*O_{G=2}^h=-Q_h(*O_{G=1}^h)=Q_h(Q(\P_h))=-Q(Q_h(\P_h))=-Q(\d C\P_h),
\label{===}
\ee
where the first equality follows from (\ref{QhO1}) and because $*$ commutes
with $Q_h$ (recall that the $*$ operation is defined by  $*(D_- Y_+ \pm D_+
Y_-) = D_- Y_+ \mp D_+ Y_-$). The second equality follows from eq.
(\ref{QQh}) and because $Q^2=0$ for all $h$ implies $Q_h^2=0$. The third
equality  follows because $Q^2=0$ for all $h$ implies  $\{Q,Q_h\}=0$, and
the last equality follows from eq. (\ref{QhP}).

To prove that $\Obt$ in (\ref{O2t}) is BRST invariant, we proceed in
precisely the same way as for $\Ob$, except that we go off-shell by
continuing $q$ rather than $h$ away from zero.  We find that
\be
Q_q(\Oa^q)=-\Obt^q
\ee
where
\be
Q_q(\P_q)=\frac12[D_-,D_+]C\P_q,\qquad Q_q(C)=0,
\ee
and $\P_q\equiv \P_{0,q}$ is a zero dimension off-shell primary superfield.
This gives (\ref{O2t}).  Again, we have explicitly checked that $\Obt$ is
not $Q$-exact and is not in the same cohomology class as $\Ob$ (see
Appendix).  Finally, it again follows that
\be
\frac12Q([D_-,D_+]C\P)=-*\Obt
\ee
and hence that $\Obt$ can be derived from a descent equation. This
completes the discussion of BRST cohomology at ghost number 2.

The conjectured BRST cohomology at $G=3$ is found analogously, either from
\be
\Oc=Q_h(Q_q(\Oa ))
\ee
or from the descent equation
\be
\frac12Q(\d C[D_-,D_+]C\P)=-*\Oc .
\ee
We have not explicitly verified that $\Oc$ is not $Q$-exact.

Next we discuss an example: the $N=2$ string in toroidal background.
In terms of $N=2$ chiral superfields
\ba
X^i(Z,\bar{Z};\th^-,\tb^-)&=&X^i_L(Z,\th^-)+X^i_R(\bar{Z},\tb^-)\nonumber\\
&=&x^i(Z,\bar{Z})+\psi_L^i(Z,\bar{Z})\th^-+\psi_R^i(Z,\bar{Z})\tb^-
+F^i(Z,\bar{Z})\th^-\tb^-\nonumber\\
Z&=&z-\th^+\th^-,
\label{X}
\ea
(where $i=s,t$ denote complex spacelike and timelike components, and bars
denote complex conjugation), the vertex operators to create a mode with
Lorentzian momentum $(p_L,p_R)$ is
\be
\P_p(X,\bar{X})=e^{i(p_L \bar{X}_L + \bar{p}_L X_L +
p_R \bar{X}_R + \bar{p}_R X_R)},
\label{Pp}
\ee
where $p_L$ and $p_R$ are complex 2-momentum, and $(p_L,p_R)$ is in the
even-self-dual lattice $\Gamma^{4,4}$ (see \cite{givshap} for more
details). The antichiral superfield $\bar{X}$ in (\ref{Pp}) is obtained
from $X$ by a complex conjugation, together with
$\th^-\leftrightarrow\th^+$. The on-shell condition is
$p_L\cdot\bar{p}_L=p_R\cdot\bar{p}_R=0$. With this condition,
$\P_p(X_L,\bar{X}_L)$ is a primary superfield with $h=q=0$, and it
generates the BRST elements given in  equations (\ref{O1})-(\ref{O3}).

In addition to these cohomology classes, there are elements of the BRST
cohomology corresponding to discrete states (that appear only at
momentum $p=0$). For a holomorphic chiral superfield $X_L$, one finds the
discrete BRST invariant operator
\be
O_{discrete}=D_+(CD_-X_L),
\label{disc1}
\ee
while for a holomorphic antichiral superfield $\bar{X}_L$ one finds the
discrete operator
\be
\bar{O}_{discrete}=D_-(CD_+\bar{X}_L).
\label{disc2}
\ee
(This can be shown by inserting $\P=X_L$ or $\P=\bar{X}_L$ into eq.
(\ref{O1}), and by using the property of (anti)chiral superfields: $D_+
X = 0$ ($D_- \bar{X} = 0$)). The discrete states corresponding to these
cohomology classes  are formed from off-shell states by taking an
appropriate limit $p\rightarrow 0$ (see ref. \cite{givshap} for details).

There are also analogous discrete states at $G=2$; these are
\be
D_+(CD_-D_+CD_-X_L),\qquad D_-(CD_+D_-CD_+\bar X_L);\label{disG=2}
\ee
relevant identities for finding these can be found in the Appendix.

\section{The twisted $N=4$ Symmetry of the $N=2$ String}
\setcounter{equation}{0}
In ref. \cite{GS} it was shown that $N=2$ string theory gives a realization
of an $N=2$ superfield extension of the topological conformal algebra. In
this section we  untwist the $N=2$ topological algebra
to get an $SU(2)\times SU(2)\times U(1)$
$N=4$ superconformal current algebra.\footnote{
A twisting of
$N=4$ superconformal field theories
to topological field theories with  $N=2$ supersymmetry was also considered
in ref. \cite{N}.}

We introduce the following $N=2$ superfields
\be
J=BC,
\label{J}
\ee
\be
G=j,
\label{G}
\ee
\be
\Gb=B,
\label{Gb}
\ee
\be
T_a=T+a\d J.
\label{Ta}
\ee
The superfield $J$ in (\ref{J}) is the ghost number current,
$G$ in (\ref{G}) is the BRST current defined in (\ref{j2}),
and $\Gb$ in (\ref{Gb}) is the superghost $B$. The stress tensor $T_a$
is a twist of the ghost+matter stress tensor $T$ defined in (\ref{TT}). The
central charge of $T$ is $c=0$,
the dimensions of $J$, $G$, $\Gb$ and $T_a$
with respect to $T$ are $h=0,0,1$ and $1$,
respectively, and their charge is $q=0$.

It is notable \cite{GS} that
the BRST charge in (\ref{Q2}) acts on $B$ and $J$ by (\ref{QO2})
\be
Q(B)=T,\qquad Q(J)=j.
\label{QB}
\ee

With the operator $\L^h$ defined in
(\ref{Lh},\ref{L}), one finds that $J,G,\Gb,T_a$ obey the algebra
\be
T_a\cdot T_a\sim \L^1 T_a+...,
\label{TaTa}
\ee
\be
T_a\cdot J \sim \L^0 J+...,
\label{TJ}
\ee
\be
T_a\cdot G\sim \L^a G+...,
\label{TG}
\ee
\be
T_a\cdot \Gb \sim \L^{1-a} \Gb+...,
\label{TGb}
\ee
\be
J\cdot G\sim -\frac{\th^+\th^-}{z}G+..., \qquad
J\cdot \Gb\sim \frac{\th^+\th^-}{z}\Gb+...,
\label{JG}
\ee
\be
G\cdot\Gb\sim \frac{\th^+\th^-}{z}(T_a-a\d J) -
(\frac{\th^+}{z}D_+-\frac{\th^-}{z}D_-)J+...,
\label{GGb}
\ee
\be
\Gb\cdot G\sim \frac{\th^+\th^-}{z}(T_a+(1-a)\d J) +
(\frac{\th^+}{z}D_+-\frac{\th^-}{z}D_-)J+... ,
\label{GbG}
\ee
\be
J\cdot J\sim G\cdot G \sim \Gb\cdot\Gb\sim 0.
\label{JJ}
\ee
In (\ref{TaTa})-(\ref{JJ}) it is understood that the first operator in the
product is at the point ($z_1,\th_1^{\pm}$) in superspace,
while the second operator in the product and the
operator on the right-hand side are at the point ($z_2,\th_2^{\pm}$).
We have
also used the shorthand notation: $\th^{\pm}\equiv \tt^{\pm}$, $z\equiv\z$.
{}From eq. (\ref{TaTa}) we learn that $T_a$ is a stress tensor with central
charge $c_a=0$ for any twist parameter $a$.
Eq. (\ref{TJ}) means that $J$ is a primary superfield of $T_a$
with $h_a(J)=0$.
Eqs. (\ref{TG}) and (\ref{TGb}) mean that $h_a(G)=a$, and $h_a(\Gb)=1-a$
with respect to $T_a$,
namely, there is a spectral flow under the twist $T\rightarrow T_a$.

We now arrive to the important observation of this section.
At $a=0$, the algebra (\ref{TaTa})-(\ref{JJ}) is the
$N=2$ superfield extension of the topological conformal algebra in
\cite{GS}, namely, the algebra generated by $T,J,B$ and $j$.
This algebra is twisted to the standard $SU(2)\times SU(2)\times U(1)$
$N=4$ superconformal algebra, with central charge $c=0$, when $a=1/2$.
Explicitly, we identify
\be
\TT =-iT_{a=1/2},\qquad \II =-\frac{1}{2}J,\qquad \AA =\frac{i}{2}(G+\Gb),
\qquad \BB =\frac{1}{2}(G-\Gb),
\label{ABIT}
\ee
where $\TT$ is a super stress tensor of conformal spin 1,
$\AA$ and $\BB$ are two spin-1/2 superfields, and $\II$ is a spin-0
superfield.
It now follows that the $N=2$ superfields
$\TT,\II,\AA,\BB$ obey the $N=4$ superconformal algebra
appearing in eq. (23) of ref. \cite{RASS} (with $c_1=c_2=c_k=\a=0$ in the
notation of \cite{RASS}), namely
\be
\TT\cdot\TT\sim -i\L^1\TT+...,
\label{balg}
\ee
\be
\TT\cdot\II\sim -i\L^0\II+...,
\ee
\be
\TT\cdot\AA\sim -i\L^{1/2}\AA+...,\qquad
\TT\cdot\BB\sim -i\L^{1/2}\BB+...,
\ee
\be
\II\cdot\AA\sim\frac{i}{2} \frac{\th^+\th^-}{z} \BB+..., \qquad
\II\cdot\BB\sim -\frac{i}{2} \frac{\th^+\th^-}{z} \AA+...,
\ee
\be
\AA\cdot\AA\sim \BB\cdot\BB\sim
-\frac{i}{2} \frac{\th^+\th^-}{z}\TT+...,
\ee
\be
\AA\cdot\BB\sim -i\L^0 \II +\frac{i}{2} \frac{\th^+\th^-}{z}\d\II+...,
\ee
\be
\II\cdot\II\sim 0.
\label{ealg}
\ee
This algebra is the so-called `large' $N=4$ superconformal algebra
\cite{N=4} with $c=\alpha=0$, where $c$ is the  central charge, and
$\alpha$ is a measure for the asymmetry between the two affine $SU(2)$
sub-algebras.\footnote{
A general $SU(2)\times SU(2) \times U(1)$ $N=4$ superconformal algebra
has a free parameter $\alpha$ measuring the asymmetry between the two
affine $SU(2)$ sub-algebras. This algebra is sometimes referred to as the
$o(4)$ $N=4$ algebra. When $\alpha=1/2$ it becomes the
$SU(2)$ $N=4$ algebra. The `large'
$N=4$ symmetry algebra of the $N=2$ string presented here
has $\alpha=0$.}
It has 16 generators (the four
components of the four superfields in (\ref{ABIT})),
which are: the spin-2 stress tensor ($\TT_1$),
4 spin-3/2 supercurrents ($\TT_{\pm},\; \AA_1$, and $\BB_1$),
7 spin-1
currents ($\TT_0,\;\AA_{\pm},\;\BB_{\pm},\;\II_1$, and $\d\II_0$)
generating the affine extension of $SU(2)\times SU(2)\times U(1)$,
and 4 spin-1/2 currents ($\AA_0,\;\BB_0$, and $\II_{\pm}$).
(Note that the spin-0 field $\II_0$ enters the right-hand side of the
algebra (\ref{balg})-(\ref{ealg}) through its derivatives only.)

The twist described in this section resembles the twist of an $N=2$
superconformal algebra to a topological conformal algebra. In the latter
case, one defines $T_a=T+a\d J$ in terms of the $N=2$ stress tensor $T$ and
$U(1)$ current $J$. It then follows that $T_a$ is a stress tensor with
central charge $c_a=c_0(1-4a^2)$, and at $a=1/2$ one recovers the $N=0$
topological conformal algebra. In this section, we have shown that an
$SU(2)\times SU(2)\times U(1)$  $N=4$ superconformal algebra can be twisted
into an $N=2$ topological conformal algebra.

Moreover, the twist of the $N=2$ critical  string algebra, regarded as a
topological superconformal algebra, into an $N=4$ superconformal algebra,
is similar to the bosonic string regarded as a topological conformal field
theory, that is `almost' a twist of an $N=2$ superconformal algebra (see
section 2).\footnote{ During the years that we have been writing up our
results, it was shown that  the $N=1$ superstring has an analogous  $N=3$
twisted supersymmetry~\cite{BLNW}.  As in the bosonic string case,  the
only requirement is that the combined supermatter and supergravity system
has a $U(1)$ supercurrent that can be used to improve the BRST current.}

In the next section we show that the algebra of {\it any} $N\geq 2$
`critical string', regarded as a topological superconformal algebra, can be
twisted into an $N+2$ superconformal algebra.\footnote{We thank Nathan
Berkovitz for pointing this out and explaining it to us.}

\section{Comments and Discussion}
\setcounter{equation}{0}

In this section we discuss the results presented in this work, and make a
few comments.

In section 4 we have shown that $N=2$ strings with total ghost+matter
central charge $c=0$ have an $N=4$ twisted symmetry.\footnote{This is not
to be confused with the {\it physical} $N=4$ supersymmetry proposed in ref.
\cite{S}.} It is remarkable that unlike the $N=2$ superconformal structure
of the $N=0$ string, and the $N=3$ superconformal structure of the $N=1$
superstring, the $N=4$ superconformal structure of the $N=2$ string does
not involve  an improvement of the BRST current. While the combined
supermatter and supergravity stress tensor is twisted, the BRST
supercurrent $j$ and ghost number supercurrent $J$ are not modified. This
occurs because the $N=2$ ghost number current is not anomalous, and hence
needs no improvement.

Iteratively, one can twist the algebra of {\it any} $N>2$  superconformally
invariant $2d$ system into an  $N+2$ superconformal algebra. One regards
the $N>2$ system as a  critical matter system, whether or not it already
has ghosts ({\it e.g.,} the $N=4$ system constructed in section 4)  and
adds (new) ghosts and antighosts.  Since for $N>2$ the ghost system has
$c=0$, the new system with the ghosts remains critical.  In particular, for
$N\geq 2$ the ghost number current $J$ is not  anomalous.\footnote{ $J\cdot
J \sim 0$ because $J$ is a sum of commuting $N=2$ currents each of which is
not anomalous.} Therefore, by twisting the stress tensor $T\rightarrow
T+\frac{1}{2}\d J$, the BRST current and the gravitational antighost can
be brought to dimensions $3/2$, and they become the two new supersymmetry
currents. The remaining lower dimensional currents of the algebra can then
be generated by the OPE's of the supercurrents.

Returning to the $N=2$ string, we note that the $N=2$ descent equations
have the novel feature that instead of involving derivatives, they involve
the operation of interchanging the real and imaginary parts of an operator
(with respect to complex conjugation defined on a Lorentz signature
worldsheet).

Finally we comment on some open issues. The relation between the ghost
number one and two (and three) BRST classes in the $N=1,2$ strings is not
completely clarified in this work. As was mentioned in the text, they are
related by a change of the ghost vacuum as well as some picture changing
operations, but the precise relation is not presented here. Furthermore,
the full cohomology at all ghost numbers has not been found.

\vskip .3in \noindent
{\bf Acknowledgements} \vskip .2in \noindent
We would like to thank Nathan Berkovitz, Alfred Shapere, Warren Siegel, Erik
Verlinde, and Barton Zwiebach for discussions.  We would like to thank the
Institute for Advanced Studies in Princeton, where most of this work was
done, and the Institute for Theoretical Physics at Santa Barbara where this
manuscript was completed.  AG would like to thank the Institute for
Theoretical Physics at Stony Brook, where part of this work was done, for
its warm hospitality. This work was supported in part by NSF grants Nos.
PHY89-04035  and PHY92-11367, as well as the John Simon Guggenheim
Foundation.

\newpage
\appendix
\section{Appendix}
\setcounter{equation}{0}

In this appendix we briefly summarize some of our computations, and
demonstrate that $\Ob$ and $\Obt$ in (\ref{O2},\ref{O2t}) are
both cohomologically nontrivial and inequivalent.

We first evaluate $Q$ on a basis of dimension zero bilinears in $C$ and
$\P$ using (\ref{jC})-(\ref{jfg}). Because $Q(\Oa )=Q(*\Oa )=0$ implies
$Q(D_\pm CD_\mp\P)=Q(CD_\pm D_\mp\P)$, we only need the following:
\ba
Q(D_+D_-C\P)&=&(CD_+D_-\d C-D_+CD_-\d C)\P\nonumber\\
&{}&+\quad D_+D_-C(D_+CD_-\P+D_-CD_+\P-C\d\P),\nonumber\\
Q(D_-D_+C\P)&=&(CD_-D_+\d C-D_-CD_+\d C)\P\nonumber\\
&{}&+\quad D_-D_+C(D_+CD_-\P+D_-CD_+\P-C\d\P),\label{QDDC}\\
Q(CD_+D_-\P)&=&C(D_+\d CD_-\P+D_+CD_-\d\P)-D_+CD_-CD_+D_-\P ,\nonumber\\
Q(CD_-D_+\P)&=&C(D_-\d CD_+\P+D_-CD_+\d\P)-D_+CD_-CD_-D_+\P ,\nonumber\\
&{}&\label{QCD}
\ea
{}From the definitions (\ref{O2}) and (\ref{O2t}) it is easy to show that
\ba
\Obt&=&\frac12*\Ob+D_-(C(D_+D_-C)D_+\P)+D_+(C(D_-D_+C)D_-\P),\nonumber\\
\frac12\Ob&=&*\Obt-D_-(C(D_+D_-C)D_+\P)+D_+(C(D_-D_+C)D_-\P).\nonumber\\
&{}&\label{OOt}
\ea
Thus $\Ob$ and $\Obt$ are cohomologically equivalent to terms without
undifferentiated $\P$'s, and hence cannot possibly be expressible in terms
of anything involving (\ref{QDDC}).  It is then straightforward to show
that no linear combination is expressible as a linear combination of
(\ref{QCD}). This proves that $\Ob$ and $\Obt$ are generically
cohomologically nontrivial and inequivalent.  Eqs. (\ref{OOt}) also are
useful for finding the $G=2$ discrete states (\ref{disG=2}).

\end{document}